\documentclass[twocolumn,prd,amsmath,amssymb, aps,superscriptaddress, nofootinbib]{revtex4-1}
\usepackage[version=4]{mhchem} 
\usepackage[normalem]{ulem}
\usepackage{amsmath}
\usepackage{amssymb}
\usepackage{amsfonts}
\usepackage{mathrsfs}
\usepackage{graphicx}
\usepackage{xcolor}
\usepackage{xfrac}
\usepackage{footmisc}
\usepackage{pifont}
\usepackage{times}
\usepackage{epstopdf}
\usepackage{enumerate}
\usepackage[hidelinks]{hyperref}
\usepackage{enumitem}
\usepackage{slashed}
\usepackage{relsize}
\usepackage{epsfig}
\usepackage{verbatim}
\usepackage{bm}
\usepackage[utf8]{inputenc}
\usepackage{graphics}
\usepackage{subfigure}
\usepackage[english]{babel}
\usepackage{orcidlink}

\definecolor{zima_blue}{HTML}{1393C1}
\hypersetup{setpagesize=false,
bookmarksnumbered=true,
bookmarksopen=true, 
colorlinks=true,
linkcolor=zima_blue,
urlcolor=zima_blue,
citecolor=zima_blue,
linktocpage=false}

\begin{document}

\title{Probing Inelastic Signatures of  Dark Matter Detection via Polarized Nucleus}

\author{Zai Yun}

\author{Junwei Sun}

\author{Bin Zhu}
\thanks{Corresponding author: \href{mailto:zhubin@mail.nankai.edu.cn}{zhubin@mail.nankai.edu.cn}}

\author{Xuewen Liu}
\thanks{Corresponding author: \href{mailto:xuewenliu@ytu.edu.cn}{xuewenliu@ytu.edu.cn}}
%\thanks{Corresponding author}
%\email{xuewenliu@ytu.edu.cn}
\affiliation{Department of Physics, Yantai University, Yantai 264005, China}

%\footnotetext[1]{Corresponding author.}

\begin{abstract}
We investigate the inelastic signatures of dark matter-nucleus interactions, explicitly focusing on the ramifications of polarization, dark matter splitting, and the Migdal effect. Direct detection experiments, crucial for testing the existence of dark matter, encounter formidable obstacles such as indomitable neutrino backgrounds and the elusive determination of dark matter spin. To overcome these challenges, we explore the potential of polarized-target dark matter scattering, examining the impact of nonvanishing mass splitting and the role of the Migdal effect in detecting light dark matter. 
Our analysis demonstrates the valuable utility of the polarized triple-differential event rate as an effective tool for studying inelastic dark matter. It enables us to investigate both angular and energy dependencies, providing valuable insights into the scattering process.

%\textcolor{red}{Our findings significantly contribute to understanding direct detection experiments, unveiling new insights into the behavior of dark matter and its inelastic nature.}
\end{abstract}

\maketitle
\section{Introduction}
\label{intro}
For several decades, the enigmatic nature of dark matter (DM), despite its dominance in the mass budget of galaxies, has puzzled the scientific community. One compelling explanation is the hypothesis that DM is composed of weakly interacting massive particles (WIMPs)~(See Ref.\cite{Jungman:1995df, Bertone:2004pz} for detail). This hypothesis arises from the intriguing coincidence between the interaction strength required for early-universe relics to possess the correct cosmological density and the electroweak interaction strength. WIMPs encompass a diverse class of DM candidates, ranging from scalar particles to fermions, with significant variations in the couplings between dark matter and ordinary matter across different models. Over the past few decades, various strategies have been proposed to detect WIMPs through space-based observations~\cite{Slatyer:2017sev}, and particle accelerators~\cite{Buchmueller:2017qhf}, with the aim of complementing direct detection techniques~\cite{MarrodanUndagoitia:2015veg, Schumann:2019eaa, Cooley:2021rws}.

Direct detection experiments, primarily focused on searching for non-relativistic DM-nucleus scattering events~\cite{Goodman:1984dc, Wasserman:1986hh}, play a crucial role in testing the WIMP hypothesis. These experiments are conducted in low-background environments located deep underground, aiming to measure the energy spectrum and direction of nuclear recoils resulting from interactions between dark matter and ordinary matter. However, challenges arise from irreducible neutrino backgrounds~\cite{Monroe:2007xp, Vergados:2008jp, Strigari:2009bq, Gutlein:2010tq, Billard:2013qya} that can mimic the dark matter signal, necessitating effective background removal techniques. Furthermore, direct detection experiments face the challenge of insensitivity to the spin of DM.

Recent advancements in direct detection research have proposed a variation known as polarized-target dark matter scattering to address these challenges~\cite{Chiang:2012ze, Franarin:2016ppr, Catena:2018uae, Jenks:2022wtj, Blaut:2022ide}. This innovative approach offers a promising solution by simultaneously addressing the background mimicry issue and providing insights into the particle nature of WIMPs, including their spin. By measuring the polarization dependence of scattering angle distributions, it becomes possible to distinguish backgrounds and shed light on crucial aspects of WIMPs upon detection. Notably, recent studies have explored the potential of direct detection experiments to discern the spin of dark matter, differentiating between fermionic and bosonic spins and higher spin states~\cite{Jenks:2022wtj}.

However, existing literature has largely overlooked the potential impact of nonvanishing mass splitting in dark matter~\cite{Tucker-Smith:2001myb, Tucker-Smith:2004mxa, Cui:2009xq, Graham:2010ca, Batell:2009vb, Arina:2009um, Izaguirre:2015zva, McCullough:2013jma, Bramante:2016rdh, Bozorgnia:2013hsa, Feldstein:2010su, DeSimone:2010tf, An:2020tcg, Hooper:2010es} on the energy deposition spectrum within the detector, particularly in relation to polarized targets. This omission is particularly pertinent in light of the constraints imposed by the Cosmic Microwave Background (CMB)~\cite{Adams:1998nr, Chen:2003gz, Slatyer:2015jla, Slatyer:2015kla}. Measurements of the CMB contradict the presence of light Dirac fermion dark matter, especially if it is capable of undergoing annihilation during the recombination era. However, the introduction of a pseudo-Dirac fermion~\cite{CarrilloGonzalez:2021lxm} allows for the circumvention of these CMB constraints, as long as only the ground state exhibits a significant abundance. Consequently, viable models involving light dark matter necessitate the presence of both a ground state and an excited state, along with a mediator in the form of a light dark photon that connects to the Standard Model.

%\textcolor{red}{
The investigation of inelastic dark matter continues to be a dynamic and thriving research field \cite{Tsai:2019buq, Filimonova:2022pkj, Berlin:2023qco, eby2023earthcatalyzed}.
In this study, we direct our attention towards the exploration of inelastic signatures with polarized target, an area that has not yet received thorough investigation. The primary objective of our research is to establish a connection between these two distinct aspects and elucidate the emergence of distinctive observables resulting from their synergistic combination.

%we focus on the investigation of inelastic signatures in the context of polarized detectors, which has not been extensively explored before. Our work aims to bridge the gap between these two aspects and shed light on the unique observables that can arise from their combination.

Furthermore, an additional aspect of inelasticity in the interaction between dark matter and nuclei arises from the Migdal effect. Although the understanding of electron emission from an atom following the sudden perturbation of its nucleus has been documented since the early 1940s~\cite{Migdal1941}, the dark matter community has recently recognized its broader implications in direct detection searches~\cite{Ibe:2017yqa, Dolan:2017xbu, Essig:2019xkx, Baxter:2019pnz, Bell:2019egg, Kahn:2021ttr, GrillidiCortona:2020owp, Liu:2020pat, Flambaum:2020xxo, Acevedo:2021kly, Bell:2021zkr, Liang:2019nnx, Liang:2020ryg, Liang:2022xbu, Wang:2021oha, Bell:2021ihi, Cox:2022ekg, Wang:2021nbf, Berghaus:2022pbu, Blanco:2022pkt, Qiao:2023pbw}. Despite the usual suppression of electromagnetic signal production compared to conventional elastic nuclear scattering rates, the domain of sub-GeV dark matter presents a unique opportunity. In this regime, the nuclear recoil energy falls below the threshold, rendering it unobservable, while the electromagnetic signal remains detectable. By capitalizing on this characteristic, numerous experiments have successfully constrained the parameter space associated with sub-GeV dark matter~\cite{XENON:2019zpr, CDEX:2019hzn, DarkSide:2022dhx, EDELWEISS:2022ktt, COSINE-100:2021poy}.

In our study, we conduct a thorough investigation of the inelastic behavior observed in dark matter-nucleus scattering. Elastic scattering prevails in the case of heavy dark matter, rendering the exclusion of the Migdal effect necessary. However, light-dark matter scenarios encounter challenges related to the threshold that impedes the detection of nucleus recoil. Therefore, incorporating the Migdal effect is crucial to augment sensitivity in these situations. 
%\textcolor{red}{
Although the Migdal effect has been studied in the direct detection literature, its exploration in the context of inelastic dark matter detection with polarized detectors is still in its infancy. We provide a comprehensive analysis of the Migdal effect in this novel framework, which allows for a more detailed understanding of the experimental implications and potential observables. 
We aim to identify notable shifts in the distribution of scattering angles and recoil energy as mass splitting varies, which can be readily observed in forthcoming direct detection experiments. 

This paper is organized as follows: Section~\ref{sec: model} introduces our general dark photon model featuring pseudo-Dirac fermion dark matter. We explicitly demonstrate the matching procedure from quarks to nucleons/nuclei. By employing pseudo-Dirac fermions, we naturally attain inelastic dark matter with a tiny mass splitting. In section~\ref{sec: compution}, we present the computational framework for both the inelastic scattering and Migdal effect. We establish that the square of the matrix element remains invariant, as its variation is proportional to $\delta_{\mathrm{DM}}/m_{\chi}$ and decouples in the limit of small mass splitting. Section~\ref{sec: result} showcases the results involving the inelastic scattering and Migdal effect, employing various benchmark values. The distinctive angular and recoil energy distribution provides insights into determining the mass splitting.

\section{Dark Photon: Bridging Quarks to Nucleons and Unveiling Inelastic Dark Matter}
\label{sec: model}
We propose a generic model that establishes a connection between a dark photon, denoted as $A^{\prime}$, and standard model quarks $f$, as well as DM, represented by $\chi$. In this model, the Lagrangian is given by
\begin{equation}
    \mathcal{L}=g_{A^{\prime}} \sum_f \bar{f}\left(x_f^V \gamma^\mu+x_f^A \gamma^\mu \gamma^5\right) f A^{\prime}_\mu +\mathcal{L}_{\chi}, 
\end{equation}
which denotes the interaction between dark photon and standard model quarks and dark matter. The product $g_{A^{\prime}}\times x_{V(A)}$ denotes the strength of the vector (axial) interaction. To account for the polarized target, we extend beyond the minimal dark photon model~\cite{Holdom:1985ag, Pospelov:2007mp, Arkani-Hamed:2008hhe} by including the axial interaction $x_A$~\cite{Baruch:2022esd}. For simplicity, we focus on the isovector form of both the vector and axial vector interactions, described by the Lagrangian:
\begin{equation}
-\mathcal{L}_{f}=g_{A^{\prime}}x_V(\bar{u} \gamma^\mu u-\bar{d} \gamma^\mu d) A_{\mu}^{\prime}+g_{A^{\prime}}x_A(\bar{u} \gamma^\mu \gamma_5 u-\bar{d} \gamma^\mu \gamma_5 d)A_{\mu}^{\prime}.
\end{equation}

In our analysis of DM-nucleus scattering, it is crucial to derive the DM-nucleon interaction from the DM-quark level interaction. The matching between quarks and nucleons is straightforward, and the nucleon coupling for the vector interaction, denoted as $\mathcal{L}_{A^{\prime} n}=c_n^V A^{\prime}_\mu \bar{n} \gamma^\mu n$, can be easily identified. Here, $n$ represents both protons and neutrons, and $c_p^V=g_{A^{\prime}}x_V$ and $c_n^V=-g_{A^{\prime}}x_V$. 
Thus, we can readily match the proton to the nucleus ($N$) in the case of a vector interaction:
\begin{equation}
\left\langle N\left(p^{\prime}\right)\left|\bar{n} \gamma^\mu n\right| N(p)\right\rangle=\overline{N}\left(\gamma^\mu F\left(q^2\right)+\frac{\sigma^{\mu \nu} q_\nu}{2 m_{N}} F_1\left(q^2\right)\right) N.
\end{equation}

In the above equation, the transfer four-momentum $q=p^{\prime}-p$ is defined by the four-momenta of the incoming and outgoing nucleus, denoted as $p$ and $p^{\prime}$, respectively. We assume equality between the form factors for the proton and neutron, denoted as $F_n(q^2) = F(q^2)$, and adopt the Helm form factor. Additionally, we consider the form factor $F_1(q^2)$, which accounts for the electric and magnetic form factors governing the magnetic moment interaction. However, in this specific process, the contribution of $F_1(q^2)$ can be neglected due to its suppression by $\mathcal{O}(q/m_{N})$, where $m_{N}$ is the mass of the nucleus.

Matching for the axial vector is a similar process. At the hadron level, the nucleon matrix element of the axial-vector current of the isovector can be decomposed into two Lorentz invariant isovector form factors: the axial form factor $G_A(Q^2)$ and the induced pseudoscalar form factor $G_P(Q^2)$, 
\begin{equation}
\begin{aligned}
&\left\langle N(p^\prime) | (\bar{u} \gamma_\mu\gamma_5 u - \bar{d} \gamma_\mu\gamma_5 d) | N(p) \right\rangle \\ 
& = \bar{N}(p^\prime) \left[ \gamma_\mu G_A(Q^2) - \frac{Q_\mu}{2m_n} G_P(Q^2) \right] \gamma_5N(p). 
\end{aligned}
\end{equation}
Here, $Q^2 = -q^2$, and the pseudoscalar form factor is commonly neglected, similar to the vector case. It is widely accepted in the literature that the effective Lagrangian, specifically formulated as Eq. (\ref{eqn:L}), adequately captures the rate calculation~\cite{Catena:2018uae}: 
\begin{equation}
-\mathcal{L}_{\mathrm{int}}=h_3\bar{N}\gamma^\mu{N}A^{\prime}_\mu+h_4\bar{N}\gamma^\mu\gamma^5{N}A^{\prime}_\mu.
\label{eqn:L}
\end{equation}

Deriving the values of $h_3$ and $h_4$ from the microscopic Lagrangian after matching is straightforward:
\begin{equation}
\begin{aligned}
h_3 &= Z g_{A^{\prime}} x_V F(q^2), \\
h_4 &= g_{A^{\prime}} x_A G_A(Q^2).
\end{aligned}
\end{equation}

In the literature, it is customary to set $h_3 = -h_4 = 1/2$ to achieve maximal parity violation. The interactions between the dark sector are described by the Lagrangian,
\begin{equation}
\mathcal{L}_\chi = \bar{\chi}\left(i \gamma^{\mu} D_{\mu}-m_{\chi}\right) \chi-\frac{\delta_{\mathrm{DM}}}{4}\left(\bar{\chi}^c \chi+\text { h.c. }\right).
\end{equation}

The interaction between dark matter and dark photon is minimally realized by the covariant derivative, $D_\mu \equiv \partial_\mu+i g_D A_\mu^{\prime}$. To capture the polarization effect, we also extend the minimal coupling $g_D$ as in
\begin{equation}
   -\mathcal{L}_{\chi}= \lambda_3 \bar{u}_{\chi}\gamma^{\mu} u_{\chi}A_{\mu}^{\prime}+\lambda_4 \bar{u}_{\chi}\gamma^{\mu}\gamma_5 u_{\chi}A_{\mu}^{\prime}.
\end{equation}

The mass splitting originates primarily from the Majorana mass term, generated through the $\chi$ and $\chi^{c}$ Yukawa couplings within the framework of the Higgs mechanism. As a result, these dark sector interactions facilitate the decomposition of the Dirac fermion into two closely degenerate Majorana mass eigenstates,
\begin{equation}
\begin{aligned}
& \chi_1=\frac{1}{\sqrt{2}}\left(\chi-\chi^c\right), m_{\chi_1}=m_\chi-\frac{\delta_{\mathrm{DM}}}{2}, \\
& \chi_2=\frac{1}{\sqrt{2}}\left(\chi+\chi^c\right), m_{\chi_2}=m_\chi+\frac{\delta_{\mathrm{DM}}}{2}, 
\end{aligned}
\end{equation}
where ${\chi}_1$ and ${\chi}_2$ denote dark matter particles with masses $m_{\chi_1}$ and $m_{\chi_2}$,  respectively,  $\delta_{\mathrm{DM}}=m_{\chi_2}-m_{\chi_1}$ stands for the actual mass splitting.

We thus investigate the inelastic aspects of the interaction between dark matter and atomic nuclei in two distinct scenarios. Firstly, we explore the impact of mass splitting ($\delta_{\text{DM}}$) on the scattering of polarized nucleus and DM when the DM has a heavy mass (100 GeV), disregarding electron ionization energy $\delta_{\text{EM}}$ (). In the second scenario, which pertains to dark matter with a smaller mass (sub-GeV), we delve into the Migdal effect. This effect emerges when the low mass of dark matter interacts with nucleons, resulting in an experimental signal that is highly insensitive and challenging to observe. However, we can detect dark matter by analyzing the electromagnetic signal emitted by the electrons surrounding the nucleons. Consequently, our primary focus in this scenario centers on the impact of mass splitting ($\delta_{\text{DM}}$) of dark matter on the scattering of DM-polarized nucleus.

\section{Computational Framework}
\label{sec: compution}

To emphasize the impact of polarization-dependent effects on physical observable and facilitate a comparison between different DM mass splitting, we introduce a fundamental quantity that solely relies on the target polarization,
\begin{equation}
\frac{d \Delta R}{d E_R d \Omega}=\frac{1}{2}\left(\frac{d^3 R(\vec{s})}{d E_R d \Omega}-\frac{d^3 R(-\vec{s})}{d E_R d \Omega}\right).
\end{equation}
Here, $\vec{s}$ represents the polarization vector of the target nuclei, as defined by $\vec{s}=2\vec{S}_N$,
where $\vec{S}_N$ is the nuclear spin operator, and $d^3R/dE_R d\Omega$ is the triple differential rate of DM-nucleus scattering events per unit detector mass,
\begin{equation}
\frac{d^3 R}{dE_Rd{\Omega}} = \frac{\rho_{\chi}}{m_{\chi}m_N} \int{d^3v vf(\vec{v})}\frac{d\sigma}{dE_R d{\Omega}}.
\label{eqn:triple}
\end{equation}

At low dark matter mass, the recoil of the nucleus falls below the detector threshold, rendering it invisible to the experiment, despite a significant cross-section between the dark matter and nucleus. The Migdal effect enables the efficient capture of ionized electrons, thereby preserving our ability to probe light-dark matter. In the case of contact interaction, the Migdal differentiation can be factorized into elastic scattering between the nucleus and dark matter, as well as the probability of ionization of electrons, 
\begin{equation}
\frac{d^4 R}{d E_{\mathrm{EM}} d E_R d \Omega}=\frac{d^3 R}{d E_R d\Omega} \times \frac{1}{2 \pi} \sum_{n, \ell} \frac{d}{d E_{\mathrm{EM}}} p_{q_e}^c\left(n, \ell \rightarrow E_e\right).
\end{equation}

This study presents the ionization probability $p_{q_e}^c$ for xenon, with specific values available in~\cite{Ibe:2017yqa}. The deposited energy spectrum is determined by summing the contributions from nuclear recoil ($E_R$) and electromagnetic energy ($E_{\mathrm{EM}}=\delta_{\mathrm{EM}}$). $E_{\mathrm{EM}}$ includes the energy of the ejected electron ($E_e$) and the atomic de-excitation energy ($E_{nl}$). Since the nuclear recoil falls below the observable threshold, it must be integrated out by considering $d\Omega$ and $dE_R$ to generate the physical differential event rate, 
\begin{equation}
\begin{aligned}
\frac{dR}{d E_{\mathrm{det}}}&=\int d\Omega \int_{E_R^{\min}}^{E_R^{\max}}d E_R\frac{d^3 R}{d E_R d\Omega} \times \frac{1}{2 \pi} \\&\sum_{n, \ell} \frac{d}{d E_{\mathrm{EM}}} p_{q_e}^c\left(n, \ell \rightarrow E_e\right) \times \delta\left(E_{\mathrm{det}}-\mathcal{L} E_R-E_{\mathrm{EM}}\right), 
\label{eqn:quadruple}
\end{aligned}
\end{equation}
where $E_{R}^{\min}$ and $E_R^{\max}$ are determined in kinematics, and $E_{\mathrm{det}}$ stands for the detected electron energy,
\begin{equation}
E_{\mathrm{det}}=\mathcal{L} E_R+E_{\mathrm{EM}}, 
\end{equation}
with $\mathcal{L}$ stands for the quenching factor.

\subsection{Kinematics}
Consider the inelastic collision between DM and a nucleus in the detector. The scattering process can be represented as ${\chi}_1N\longrightarrow{\chi}_2 N$ and $N$ represents the Xenon nucleus with a mass of $m_N = 122~\mathrm{GeV}$. 

\begin{figure}[htbp]
    \centerline{ 
    \includegraphics[width=9cm]{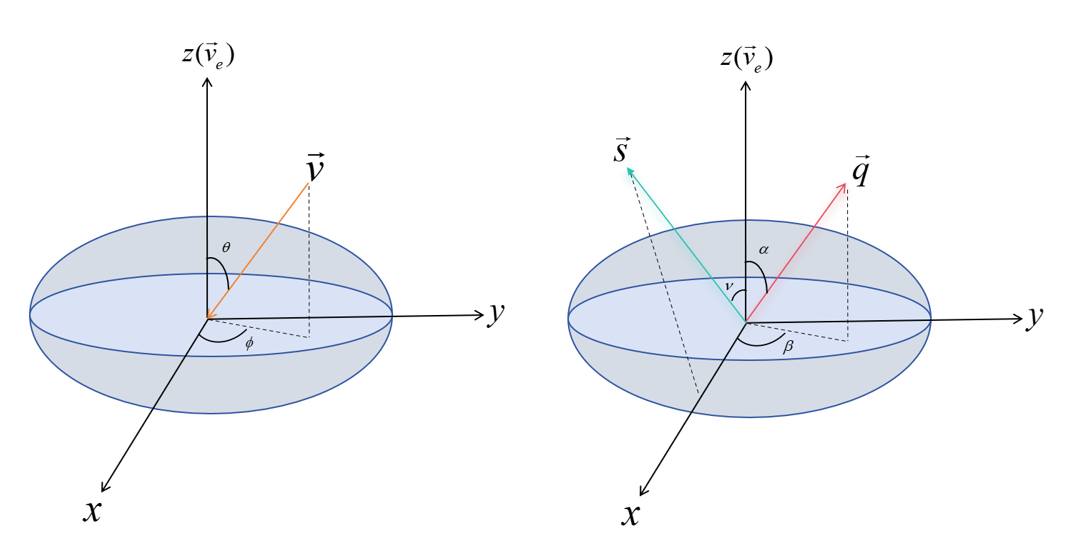}}
    \caption{A three-dimensional coordinate system with the earth velocity direction in the Z-axis direction is constructed to calculate the triple differential rate of dark matter scattering and nucleon scattering. Among them, the direction of the orange arrow is the direction of the incident velocity of the dark matter, the direction of the indigo blue arrow is the direction of the polarization of the nucleon, and the red is the direction of the recoil of the nucleon.}
    \label{Fig.1}
    \end{figure}

In a three-dimensional coordinate system, let $\vec{v}$ represents the incoming velocity of the DM particle at the detector, and $\vec{v}^\prime$ represents the outgoing velocity of the DM. This vector $\vec{v}$ makes an angle $\phi$ with the x-axis and an angle $\theta$ with the z-axis, aligning with the unit vector $\hat{v}=(\sin \theta \cos \phi, \sin \theta \sin \phi, \cos \theta)$. As for the direction of the nuclear recoil, $\vec{q}$ makes an angle $\beta$ with the x-axis and an angle $\alpha$ with the z-axis in the same three-dimensional coordinate system. The unit vector of transferred momentum is denoted as $\hat{q}=(\sin \alpha \cos \beta, \sin \alpha \sin \beta, \cos \alpha)$. To facilitate the discussion and explicitly showcase the angle dependence, let us begin by considering a specific scenario where $\alpha=\theta={\pi}/{2}$ and $\phi=0$. This simplification enables us to treat the inelastic collision between dark matter and nucleons as occurring on a two-dimensional plane. However, it is crucial to note that the entire scattering process can still be reconstructed in a three-dimensional plane. The conservation of energy and momentum leads to the following relationships:
\begin {equation}
\begin {aligned}
    m_{\chi_1}+\frac{1}{2}m_{\chi_1}v^2&=m_{\chi_2}+\frac{1}{2}m_{\chi_2}v^{\prime 2}+\frac{q^2}{2m_N},
    \\ m_{\chi_1} v &=m_{\chi_2}v^{\prime}\cos{\phi_1}+q\cos{\beta},
    \\ 0&=q\sin{\beta}-m_2v^{\prime}\sin{\phi_1}, 
    \label{eqn:conservation}
    \end{aligned}
\end{equation}
where $\phi_1$ stands for the out-going dark matter direction. By applying energy and momentum conservation, we determine
\begin{equation}
q=2\mu v \cos{\beta}-\frac{\Delta}{\cos{\beta}v}.
\end{equation}
In this equation, $\Delta=\delta_{\mathrm{DM}}+\delta_{\mathrm{EM}}$, in the inelastic collision between dark matter and the nucleus. The DM-target reduced mass $\mu=\frac{m_Nm_{\chi}}{m_N+m_{\chi}}$. 
The triple differential scattering cross-section $d\sigma/dE_R d\Omega$ for the scattering of dark matter on a polarized target can be expressed within the energy-momentum conservation,
\begin{equation}
\frac{d{\sigma}}{dE_Rd\Omega}=\frac{d{\sigma}}{2{\pi}dE_R}\delta\left(\cos{\beta}-\frac{q}{2{\mu}v}-\frac{\Delta}{qv}\right).
\label{eqn: cross-section}
\end{equation}
Recall that equation ~(\ref{eqn: cross-section}) simplify $\delta$ function, we have
\begin{equation}
\begin{aligned}
    \frac{d{\sigma}}{2{\pi}dE_R}{\delta}\left(\cos{\beta}-\frac{q}{2{\mu}v}-\frac{\Delta}{qv}\right)
    &=\frac{d{\sigma}v}{2{\pi}dE_R}{\delta}\left(\vec{v}\cdot \hat{q}-\frac{q}{2{\mu}}-\frac{\Delta}{q}\right)
    \\&=\frac{d{\sigma}v}{2{\pi}dE_R}\frac{{\delta}(v-\bar{v})}{|\hat{v}\cdot\hat{q}|}, 
    \label{eqn: d}
    \end{aligned}
\end{equation}
where $\bar{v} = q/(2{\mu}(\hat{v}\cdot\hat{q}))+\Delta/q(\hat{v}\cdot\hat{q})$ and $\hat{v}$ is a unit vector in the direction of the incoming DM velocity, $\vec{v}=v\hat{v}$. 
The velocity distribution of DM can significantly affect the rates of direct detection. The Maxwell-Boltzmann distribution is commonly used as a simple analytic approximation for the velocity distribution of DM. Within the frame of the Galaxy, this velocity distribution is
\begin{equation}
f(\vec{v})=\frac{1}{\mathcal{N}}e^{-(\vec{v}+\vec{v_e})^2/{v_0}^2}\Theta({v_{\mathrm{esc}}-|\vec{v}+\vec{v_e}|}), 
\end{equation}
with a galactic escape velocity of $v_{\mathrm{esc}}=544\mathrm{km/s}$, the distribution is cut off at the local escape
speed, and a most probable speed given by the circular speed of the local standard of rest of $v_0=220,\mathrm{km/s}$. The Earth's velocity in the galactic rest frame, $\vec{v_e}$, is $232,\mathrm{km/s}$. $\Theta$ is the Heaviside step function, and $\mathcal{N}$ is a normalization constant, 
\begin{equation}
\mathcal{N}=\pi{v_0}^2\left(\sqrt{\pi}v_0\mathrm{Erf}\left(\frac{v_{\mathrm{esc}}}{v_0}\right)-2v_{\mathrm{esc}}e^{-{v_{\mathrm{esc}}}^2/{v_0}^2}\right).
\end{equation}
The velocity range considered is $v\in[v_{\min},v_{\max}]$, where $v_{\min}=\sqrt{{m_NE_R}/{2\mu^2}}$ and $v_{\max}=v_{\text{esc}}$, with $m_N$ denoting the mass of the target nucleus, $E_R$ the recoil energy, and $\mu$ the reduced mass of the dark matter-nucleus system. The problem requires $v_{\mathrm{esc}}-|\vec{v}+\vec{v_e}|>0$, which implies $\cos\theta<({v_{\mathrm{esc}}^2-v^2-v_e^2})/{2 v v_e}$. The range of $\cos\theta$ is therefore $[-1,{(v_{\mathrm{esc}}^2-v^2-v_e^2)}/{2 v v_e}]$. However, this range over-counts the available parameter space since $|\vec{v}+\vec{v_e}|$ is smaller than $v_{\mathrm{esc}}$ by definition. Consequently, $\cos\theta$ can cover the entire parameter space $\cos\theta\in[-1,1]$. The maximum value of $|\vec{v}+\vec{v_e}|$ occurs when $\vec{v}$ is parallel to $\vec{v_e}$. Thus, the corresponding ranges of $v$ and $\cos\theta$ are $v\in[v_{\min},v_{\mathrm{esc}}-v_e]$ and $\cos\theta\in[-1,1]$. Another phase space exists where $\cos\theta\in[-1, (v_{\mathrm{esc}}^2-v^2-v_e^2)/{2 v v_e}]$ and $v\in[v_{\mathrm{esc}}-v_e,v_{\mathrm{esc}}+v_e]$. Thus the velocity integral becomes,
\begin{equation}
\begin{aligned}
\int d^3v &= \int_{v_{\min}}^{v_{\mathrm{esc}}-v_e}dv v^2\int_{-1}^{+1}d\cos\theta\int_0^{2\pi}d\phi \\
&+\int_{v_{\mathrm{esc}}-v_e}^{v_{\mathrm{esc}}+v_e}dv v^2\int_{-1}^{\frac{{v_{\mathrm{esc}}}^2-v^2-{v_e}^2}{2vv_e}}d\cos\theta\int_0^{2\pi}d\phi.
\label{eqn: velocity integral}
\end{aligned}
\end{equation}

Performing the velocity integral defined in equation~(\ref{eqn: velocity integral}) while respecting energy-momentum conservation as described in equation~(\ref{eqn: d}) determines the triple differential event rate, which is the cornerstone of the calculation: 
\begin{equation}
\begin{aligned}
\frac{d^3 R}{dE_R d{\Omega}}=\frac{\rho_\chi}{64\pi^2m_{\chi}^3m_N^2\mathcal{N}}\sum\limits_{l=1}^2\int_{-1}^{+1}d\cos\theta\int_0^{2\pi}d\phi
\\\frac{\bar{v}^{2}}{|\hat{v}\cdot\hat{q}|}e^{-(\bar{v}^{2}+v_e^2+2\bar{v}^{\prime}v_e\cos{\theta})/v_0^2}|\overline{M}|^2\Theta_l.
\label{eqn: R}
\end{aligned}
 \end{equation}

In Migdal scattering, the incoming and outgoing states exhibit slight deviations from the elastic scattering process involving a DM particle, an ionized atom, and an unbound electron. The incoming DM is treated as a plane wave, representing an energy eigenstate and a momentum eigenstate. Similarly, the incoming atom, initially at rest in the lab frame, is considered both an energy eigenstate and a momentum eigenstate with respect to the total momentum of the atom. Consequently, the entire atom experiences recoil with a velocity $v_A$ and possesses momentum $p_A = m_A v_A$, where $m_A = m_N$ given the negligible electron mass. In the case where the atom is regarded as a composite system comprising electrons and a nucleus with multiple internal energy levels, conservation laws dictate the energy and momentum in DM-atom interactions. For DM with mass $m_\chi$, an incoming velocity $v$, and an outgoing momentum $p_{\chi}^\prime$, the conservation of energy and momentum results in the following relationship 
\begin{equation}
E_R=\frac{\mu^2}{m_A} v^2\left[1-\frac{\Delta}{\mu v^2}-\sqrt{1-\frac{2 \Delta}{\mu v^2}} \cos \theta_{\mathrm{cm}}\right], 
\label{eqn: ER}
\end{equation}
with $E_{R}^{\min}$ and $E_R^{\max}$ corresponding to $\cos\theta_{\mathrm{cm}}=\pm 1$.

\subsection{Dynamics}
The square of the matrix element, $|M|^2$, encapsulates the underlying physics of the scattering process and is contingent on the particular DM model being investigated. In this study, we only consider fermionic dark matter as the object of study. To describe the scattering amplitude, particularly in situations that involve heavy mediators with a significantly larger mass ($m_{A^{\prime}}$) compared to the momentum transfer ($q$), we can utilize the Lagrangian presented in Equation (\ref{eqn:L}) along with its corresponding Feynman rule, 
\begin{equation}
   \begin{aligned}
     iM= \frac{-i}{{m_{A^{\prime}}}^2}\bar{u}_{\chi_2}(p^{\prime},s^{\prime})\gamma^\mu(\lambda_3+\lambda_4\gamma_5)u_{\chi_1}(p,s)&\\\bar{u}_{N}(k^{\prime},r^{\prime})\gamma_\mu(h_3+h_4\gamma_5)u_{N}(k,r).
     \label  {eqn:im}
   \end{aligned} 
\end{equation}
In this context, the symbols $\bar{u}_{\chi_2}$, $u_{\chi_1}$, $\bar{u}_{N}$, and $u_{N}$ represent the solutions derived from the Dirac equation. The variables $s$ and $s^{\prime}$ ($r$ and $r^{\prime}$) correspond to the initial and final spins of the DM particle and the target nucleus, respectively. When considering the non-relativistic limit, the spinor bilinears can be expressed as follows~\cite{Cirelli:2013ufw},

\begin {equation}
\begin {aligned}
\bar{u}_{\chi_2}(p^{\prime},s^{\prime})\gamma^{\mu}u_{\chi_1}(p,s) = {}&\left(
\begin{array}{c}
(2m_{\chi}+\Delta)\delta^{s^{\prime}s}\\
\vec{P}\delta^{s^{\prime}s}-2i\vec{q}\times\vec{S}_{\chi}^{s^{\prime}s}\\
\end{array} \right).
\end {aligned}
\end {equation}
\begin {equation}
\begin {aligned}
\bar{u}_{\chi_2}(p^{\prime},s^{\prime})\gamma^{\mu}\gamma_5u_{\chi_1}(p,s) = {}&\left(
\begin{array}{c}
2\vec{P}\cdot\vec{S}_{\chi}^{s^{\prime}s}\\
(4m_{\chi}+2\Delta)\vec{S}_{\chi}^{s^{\prime}s}
\end{array} \right) .
\end {aligned}
\end {equation}
 \begin {equation}
\begin {aligned}
\bar{u}_N(k^{\prime},r^{\prime})\gamma^{\mu}u_N(k,r) = {}&\left(
\begin{array}{c}
2m_{N}\delta^{r^{\prime}r}\\
-\vec{K}\delta^{r^{\prime}r}-2i\vec{q}\times\vec{S}_{N}^{r^{\prime}r}\\
\end{array} \right) .
\end {aligned}
\end {equation}

\begin {equation}
\begin {aligned}
 \bar{u}_N(k^{\prime},r^{\prime})\gamma^{\mu}\gamma_5u_N(k,r) ={} & \left(
\begin{array}{c}
2\vec{K}\cdot\vec{S}_{N}^{r^{\prime}r}\\
-4m_{N}\vec{S}_{N}^{r^{\prime}r}\\
\end{array} \right) .
\end {aligned}
\end {equation}
It should be noted that for the process ${\chi}_1N\longrightarrow{\chi}_2 N$ , since we only consider that the mass of the dark matter changes in the initial state and the final state, while the mass of the nucleon remains unchanged in the initial state and the final state, the bilinear spinor of the dark matter above has mass splitting, while the bilinear spinor of the nucleon does not have mass splitting. 
Here we define $\vec{P}=\vec{p}+\vec{p}^{\prime}$, $\vec{K}=\vec{k}+\vec{k}^{\prime}$, $\vec{S}_{N}^{r^{\prime}r}=\xi^{r^{\prime}\dagger}(\vec{\sigma_N}/2)\xi^{r}$, $\vec{S}_{\chi}^{s^{\prime}s}=\xi^{s^{\prime}\dagger}(\vec{\sigma_{\chi}}/2)\xi^{s}$, $\xi^{s^{\prime}\dagger}\xi^{s}=\delta^{s^{\prime}s}$, $\xi^{r^{\prime}\dagger}\xi^{r}=\delta^{r^{\prime}r}$, and the two component spinor $\xi$ be normalized as usual. We thus obtain the amplitude from equation~(\ref{eqn:im})
\begin{equation}
   \begin{aligned}
iM={} &\frac{-i}{{m_{A^{\prime}}}^2}\{ \lambda_3h_3[4m_Nm_{\chi}\delta^{s^{\prime}s} \delta^{r^{\prime}r}+2m_N\Delta\delta^{s^{\prime}s} \delta^{r^{\prime}r}]    
\\& +\lambda_3h_4[-8m_Nm_{\chi}\delta^{s^{\prime}s}\vec{v}^{\perp}\cdot\vec{S}_{N}^{r^{\prime}r}
\\&+8im_N \vec{S}_{\chi}^{s^{\prime}s}\cdot(\vec{S}_{N}^{r^{\prime}r}\times\vec{q})+2\Delta\delta^{s^{\prime}s}\vec{K}\cdot\vec{S}_{N}^{r^{\prime}r}]
\\&
+\lambda_4h_3[8m_Nm_{\chi}\delta^{r^{\prime}r}\vec{v}^{\perp}\cdot\vec{S}_{\chi}^{s^{\prime}s}
\\&+4i(2m_{\chi}+\Delta) \vec{S}_{\chi}^{s^{\prime}s}\cdot(\vec{S}_{N}^{r^{\prime}r}\times\vec{q})-2\Delta\delta^{r^{\prime}r}\vec{K}\cdot\vec{S}_{\chi}^{s^{\prime}s}]
\\&
+\lambda_4h_4[-16m_Nm_{\chi}\vec{S}_{\chi}^{s^{\prime}s}\cdot\vec{S}_{N}^{r^{\prime}r}-8m_N\Delta\vec{S}_{\chi}^{s^{\prime}s}\cdot\vec{S}_{N}^{r^{\prime}r}]
\}, 
 \end{aligned}
\end{equation}
where the transverse
relative velocity vector 
\begin{equation}
2\vec{v}^{\perp}=\vec{v}+\vec{v}^{\prime}+\frac{m_\chi}{m_N}\left(\vec{v}^{\prime}-\vec{v}\right).
\end{equation}

We proceed to calculate $|\overline{M}|^2$, which represents the squared amplitude for DM-nucleus scattering. This calculation involves summing over the final spin states of both the DM and the nucleus, and averaging over the initial spin configurations of the DM. Finally, we apply the following spin summation rules.

\begin{equation}
\begin{aligned}
&|\overline{M}|^2=\frac{1}{2}\sum\limits_{ss^{\prime}}\sum\limits_{r^{\prime}}|M|^2,\\
&\sum\limits_{r^{\prime}}\vec{S}_{N}^{r^{\prime}r}\times\vec{S}_{N}^{rr^{\prime}} =\frac{i}{2}\vec{s},\\
&\sum\limits_{r^{\prime}}\vec{S}_{N}^{r^{\prime}r}\cdot\vec{S}_{N}^{rr^{\prime}}=\frac{3}{4},\\
&\sum\limits_{ss^{\prime}}(\vec{a}\cdot\vec{S}_{\chi}^{ss^{\prime}})(\vec{b}\cdot\vec{S}_{\chi}^{s^{\prime}s})=\frac{1}{2}\vec{a}\cdot\vec{b}.
\end{aligned}
\end{equation}
In this analysis, we only consider terms that are linear or lower order in $\vec{q}$ and $\vec{v}^{\perp}$. Consequently, we can neglect terms such as $\vec{q} \vec{v}^{\perp}$, $\vec{v}^{2\perp}$, and $\vec{q}^2$. It is found that $|\overline{M}|^2$ can be simplified to yield (see Appendix \ref{app:validation} for details),

\begin{equation}\label{3.19}
\begin{aligned}
&|\overline{M}|^2= \frac{16m_N^2m_{\chi}^2}{{m_{A^{\prime}}}^4}\{A-B\vec{v}\cdot \vec{s}-C\vec{v}'\cdot \vec{s} \}, 
\end{aligned}
\end{equation}
with
\begin{equation}\label{3.20}
\begin{aligned}
 &A=\lambda_3^2h_3^2(1+\frac{\Delta}{m_{\chi}})+3\lambda_4^2h_4^2(1+\frac{\Delta}{m_{\chi}}), 
 \\ 
 &B=\lambda_3^2h_3h_4(1-\frac{m_{\chi}}{m_N}+\frac{\Delta}{4m_{\chi}}(1-\frac{m_{\chi}}{m_N})-\frac{\Delta}{4m_N})\\
 &+\lambda_4^2h_3h_4(2-3(1-\frac{m_{\chi}}{m_N})-\frac{\Delta}{2m_{\chi}}(1-\frac{m_{\chi}}{m_N})+\frac{\Delta}{2m_N})\\ &+2\lambda_{3}\lambda_{4} h_{4}^{2}(1+\frac{\Delta}{2m_{\chi}}), \\ &C=\lambda_3^2h_3h_4(1+\frac{m_{\chi}}{m_N}+\frac{\Delta}{4m_{\chi}}(1+\frac{m_{\chi}}{m_N})+\frac{\Delta}{4m_N})\\&+\lambda_4^2h_3h_4(2-3(1+\frac{m_{\chi}}{m_N})-\frac{\Delta}{2m_{\chi}}(1+\frac{m_{\chi}}{m_N})-\frac{\Delta}{2m_N})\\ 
 &-2\lambda_3\lambda_4h_4^2(1+\frac{\Delta}{2m_{\chi}}).
\end{aligned}
\end{equation}
 
It is evident that the squared matrix element remains invariant under the limit of small mass splitting $\Delta/m_{\chi}\rightarrow 0$. Therefore, the deviation from elastic scattering is based on the kinematics, which will be explicitly presented in the next section.

\section{Results}
\label{sec: result}

In the context of scattering between heavy dark matter and polarized nuclei, with a benchmark value of $m_{\chi}=100~\mathrm{GeV}$, the squared matrix element is expressed as a function of $\delta_{\mathrm{DM}}/m_{\chi}$. This implies that a small mass splitting during the scattering of dark matter with nucleons does not affect the calculation of the scattering amplitude. Therefore, the presence of mass splitting in dark matter only affects the kinematics, specifically the phase space integral, while leaving the elastic scattering amplitude unaffected. We analyze the magnitude of the scattering rate for heavy dark matter in three different directions defined by the angles $\alpha$, $\beta$, and the nucleon polarization angle $\nu$ (the unit of all angles is radian). By considering the conservation of energy equation (\ref{eqn:conservation}) in the center-of-mass frame and the recoil energy of the nucleus, we find that the mass splitting falls within the range of $\delta_{\mathrm{DM}} \leq {{\mu}v^2}/{2}$ for heavy dark matter-polarized nucleus scattering. Taking $m_{\chi}=100~\mathrm{GeV}$, $m_N=122 \mathrm{GeV}$, and $v\sim10^{-3}$ as benchmark values, the resulting mass splitting is $\delta_{\mathrm{DM}} \leq 50~\mathrm{keV}$. For negative mass splitting, where dark matter scatters from heavier to lighter dark matter, the constraint on $\delta_{\mathrm{DM}}$ arises from the condition that the minimum velocity $v_{\min}$ must not exceed the escape velocity. Therefore, we focus on mass splittings within the range of $\delta_{\mathrm{DM}} \in [-40, 40]$ for the figures presented. Similarly, based on the relationship $E_{R}^{\max}=\frac{1}{2}m_Nv_{\max}^2$, we set $E_{R}^{\max}$ to be $100 ~\mathrm{keV}$. Thus, our primary focus is on the recoil energy of the nucleus, which ranges from $E_{R} = 0$ to $100~\mathrm{keV}$ in the depicted figures.

\begin{figure}[htbp]
    \centerline{ 
    \includegraphics[width=9cm]{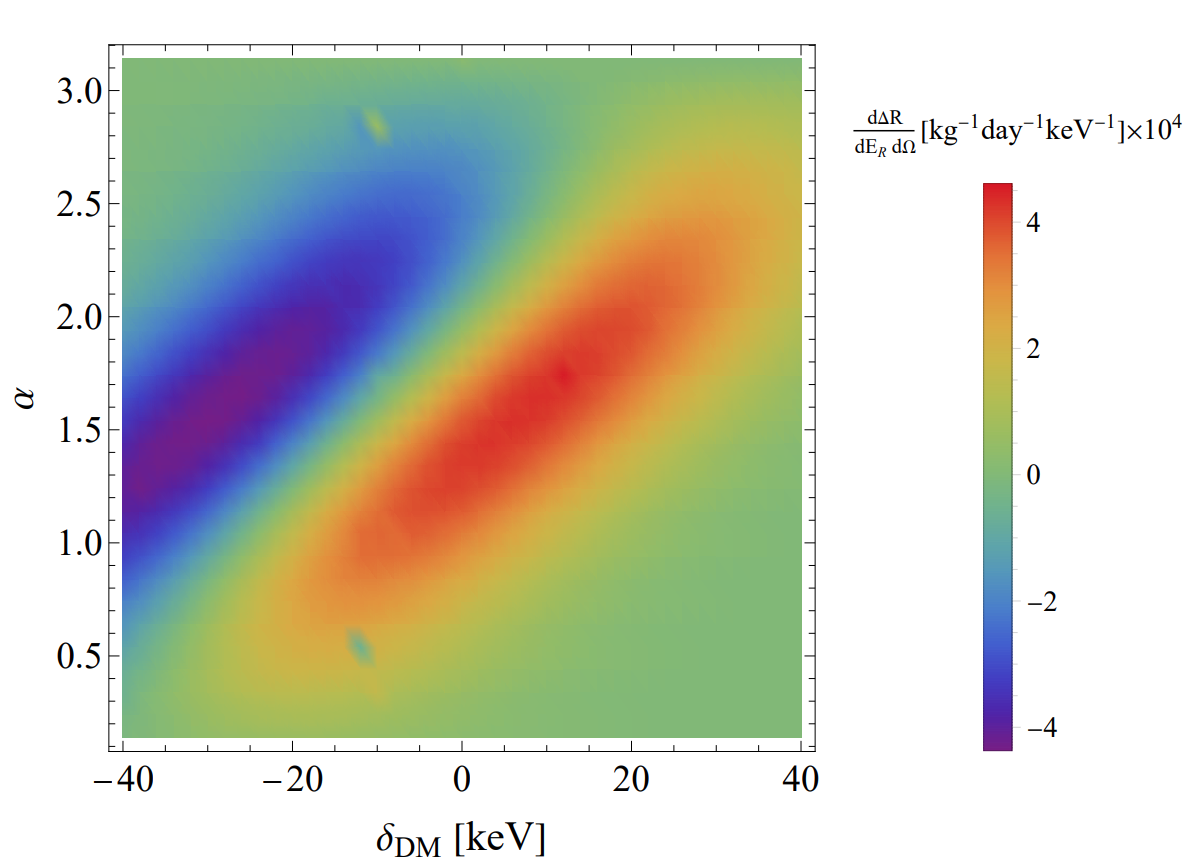}}
    \caption{The differential rate of scattering events, specifically the purely polarization-dependent component, is plotted against the polar recoil angle $\alpha$ for various mass splittings. The recoil energy $E_{R}$ is fixed at 5 $\mathrm{keV}$, and the polarization angle is set to $\nu = \pi/2$ with $\beta = 0$.}
    \label{Fig.2}
    \end{figure}

Figure \ref{Fig.2} depicts the increasing trend of the maximum scattering rate's position along the direction angle $\alpha$ with an increase in mass splitting. This trend occurs due to the correlation between an increase in $\delta_{\mathrm{DM}}$ and a corresponding increase in $\bar{v}$. As a result, the integral value increases across the entire phase space. The azimuth angle $\alpha$ is determined by $\hat{v}\cdot\hat{q}$, and in order to maintain the same scattering rate, $\hat{v}\cdot\hat{q}$ must increase as the phase space integral increases. Consequently, as $\delta_{\mathrm{DM}}$ increases, the region of the $\alpha$ angle with the local maximum scattering rate shifts towards larger values. However, as $\delta_{\mathrm{DM}}$ increases, the phase space integral expands, while the minimum value of $v_{\min}$ also increases, resulting in a contraction of the integral range. Consequently, after reaching the local maximum, the scattering rate decreases as $\delta_{\mathrm{DM}}$ increases. Conversely, when $\delta_{\mathrm{DM}} \textless 0$, the local maximum transitions to a negative value, indicating the predominance of the spin-flip component. Therefore, we can utilize the trend in $\alpha$ to differentiate between the mass splitting case and elastic scattering.

\begin{figure}[htbp]
    \centerline{ \includegraphics[width=9cm]{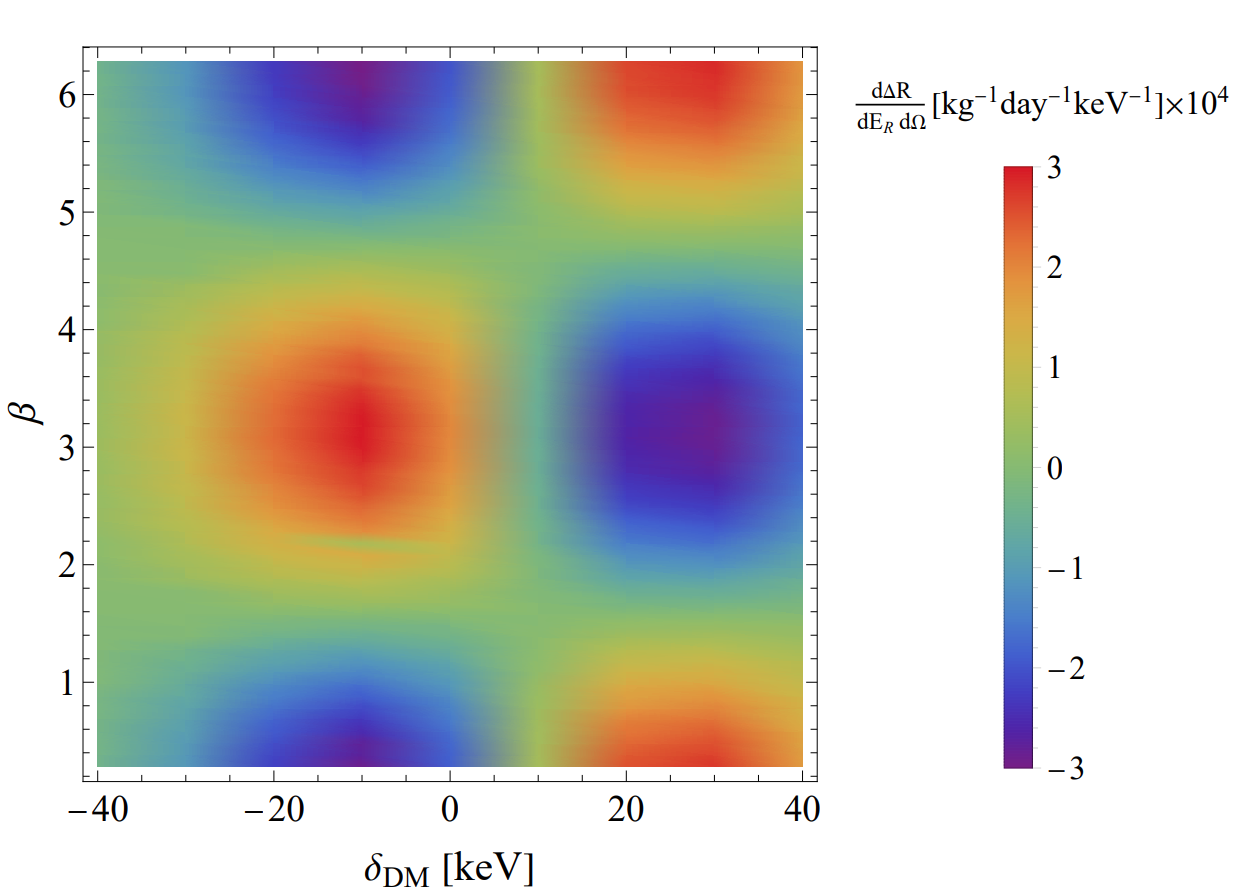}}
    \caption{The polarization-dependent component of the differential scattering event rate is plotted against the polar recoil angle $\beta$ for various mass splittings. The recoil energy  $E_R$ is fixed at $5~\mathrm{keV}$, and the polarization angle is set to $\nu = {\pi}/2$, while the polar angle is fixed at $\alpha = 3\pi/4$.}
    \label{Fig.3}
    \end{figure}

Regarding the distribution on the $\delta_{\mathrm{DM}}-\beta$ plane shown in Figure~\ref{Fig.3}, our focus is on the phase space integral, which incorporates $\beta$ through $\hat{v}\cdot\hat{q}$, enabling us to observe periodic variations in the angular distribution. Notable instances of dark matter and nucleon scattering take place at $\beta=0,\pi/2,\pi$, regardless of the positive or negative value of the mass splitting. However, the positions of the maxima interchange for positive and negative $\delta_{\mathrm{DM}}$ values.

\begin{figure}[htbp]
    \centerline{ \includegraphics[width=9cm]{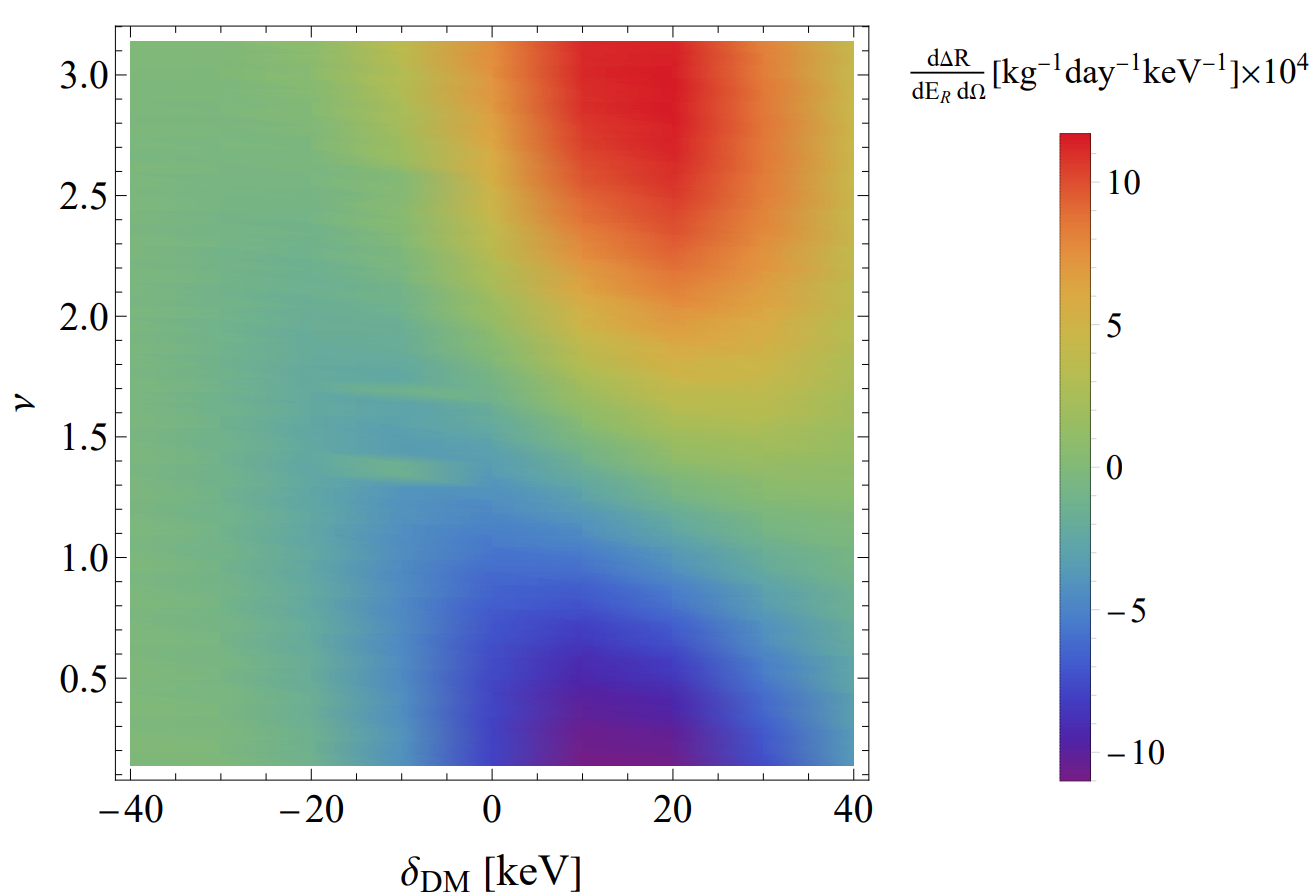}}
    \caption{The polarization-dependent component of the differential scattering event rate is plotted against the polar recoil angle $\nu$ for various mass splittings. The recoil energy $E_{R}$ is fixed at 5 $\mathrm{keV}$, and the polarization angle is fixed at $\beta = 0$, while the polar angle is fixed at $\alpha = 3\pi/4$.}
    \label{Fig.4}
    \end{figure}

The distribution on the $\delta_{\mathrm{DM}}-\nu$ plane depicted in Figure~\ref{Fig.4} reveals a substantial number of events when the mass splitting $\delta_{\mathrm{DM}}$ is positive, while the number of cases is negligible for negative mass splitting. As we have selected the z-axis direction of our coordinate system to align with the velocity $\vec{v_e}$ of the Earth in the galactic rest frame, it results in the incident velocity $\vec{v}$ of the dark matter being initially opposite to the polarization direction $\vec{s}$. Consequently, we observe a local maximum value of $\nu$ in the opposite direction within the range of $\nu \in [0, \pi/2]$. Furthermore, as the angle $\nu$ increases, the scattering rate decreases and reaches $0$ at $\nu = \pi/2$. When $\nu$ lies within the range of $\nu \in [0, \pi/2]$, $\vec{s}$ aligns with the dark matter incident velocity $\vec{v}$, resulting in a local maximum in the correct direction. Equation (\ref{eqn: R}) reveals the presence of $\bar{v}$, $\vec{v} \cdot \vec{s}$, which encompass the variables in Figure $\delta_{\mathrm{DM}}-\nu$, namely the mass splitting $\delta_{\mathrm{DM}}$  and the polarization angle $\nu$. As for the remaining values in the equation, we assign specific parameters to simplify equation (\ref{eqn: R}) for calculating the scattering rate, resulting in a simplified function of the form $(0.00035 + x/35000)^3 \exp[-(0.00035 + x/35000)^2]$. Here, the variable $x$ corresponds to the variable $\delta_{\rm DM}$ in the equation. Since our mass splitting value, $\delta_{\mathrm{DM}}$, falls within the range of $[-40, 40]~\mathrm{keV}$, our figure of $\delta_{\mathrm{DM}}-\nu$ only represents the number of cases where $\delta_{\mathrm{DM}}$ is positive.

\begin{figure*}[htbp]
    \centering 
    \includegraphics[width=19cm]{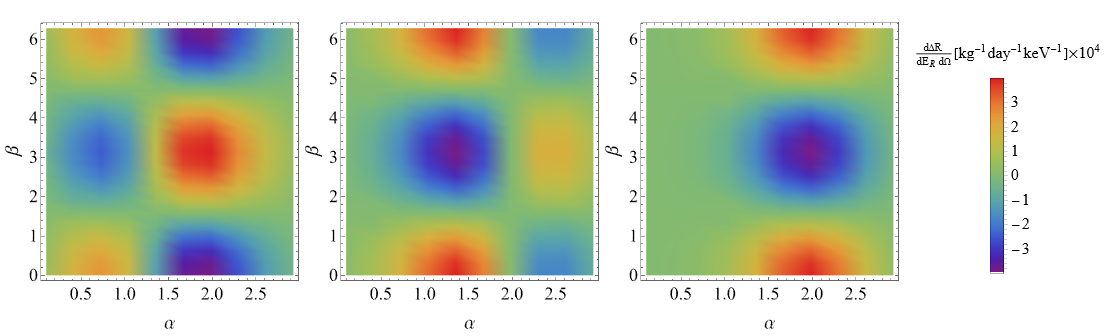}
    \caption{Purely polarisation dependent part of the differential rate of scattering events plotted against the polar recoil angle $\alpha$  with the azimuthal angle $\beta$. The recoil energy $E_{R}$ is set to 5 KeV and the polarisation angle is fixed to $\nu = \pi/2$, the left image shows the mass split as $\delta= -20 \mathrm{keV}$, the middle image shows the mass split as $\delta= 0 \mathrm{keV}$, and the right image shows the mass split as $\delta= 20\mathrm{keV}$ .}
    \label{Fig.5}
    \end{figure*}

To further investigate the effect of the mass splitting, we generate a figure of $\alpha-\beta$ in Figure~\ref{Fig.5} while holding $E_R$ constant at $5~\mathrm{keV}$ and $\nu$ at $\pi/2$. Furthermore, notable occurrences can be observed at $\beta=0,\pi/2,\pi$, and when $\delta_{\mathrm{DM}}$ is positive or negative, the positions of the positive maximum and negative maximum values are reversed, aligning precisely with the $\beta-\delta_{\mathrm{DM}}$ diagram. Similarly, we note that for $\alpha>\pi/2$, a local maximum value is observed when $\delta_{\mathrm{DM}}$ is negative, while for $\alpha=\pi/2$, a local maximum value is observed when $\delta_{\mathrm{DM}}$ is zero, and for $\alpha>\pi/2$, a local maximum value is observed when $\delta_{\mathrm{DM}}$ is positive. Therefore, in the presence of a mass splitting, the local maximum value is accompanied by an increase in the angle $\alpha$, which is reflected in the $\alpha-\delta_{\mathrm{DM}}$ plot. 

\begin{figure}[!htbp]
\centering
\includegraphics[width=9cm]{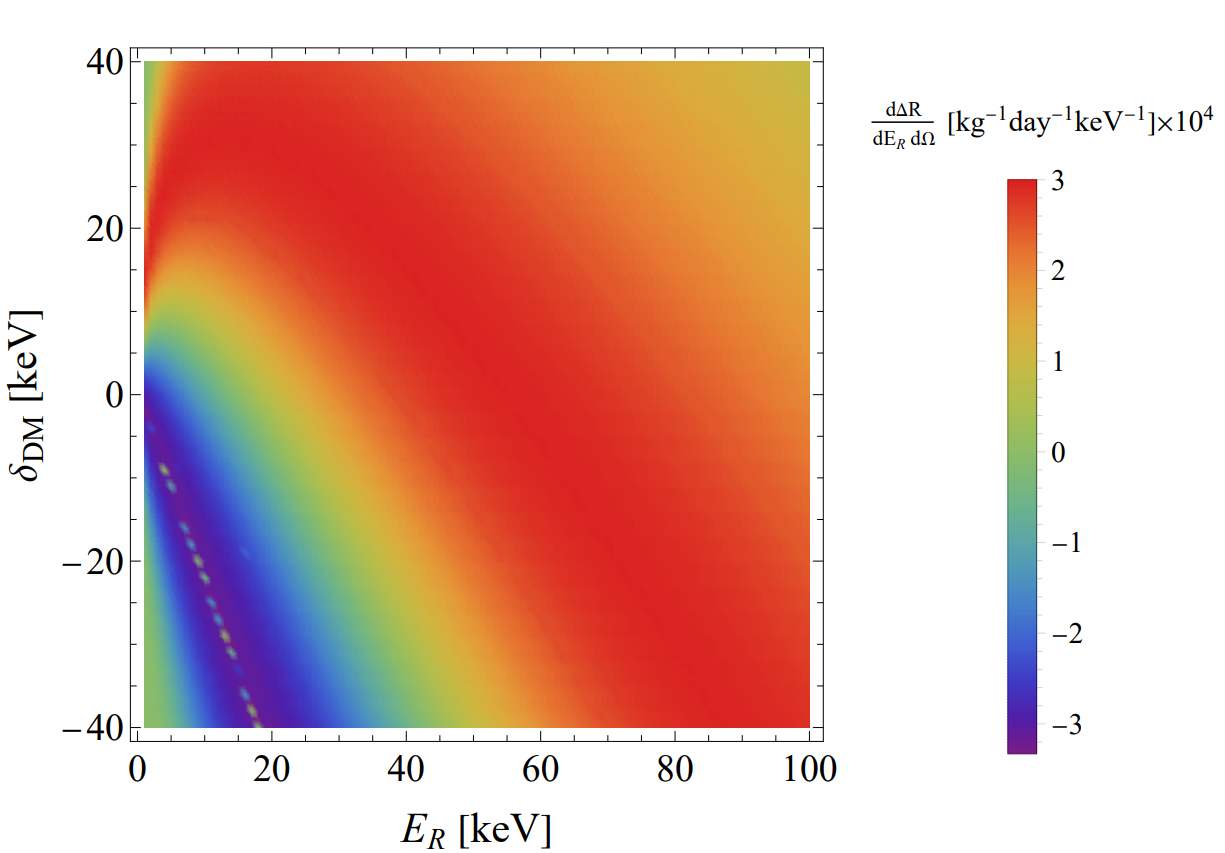}
\caption{Pure polarization dependent part of the differential rate of scattering events corresponding to energy and mass splitting. The polarization Angle is set to $\alpha = 3\pi/4, \beta=0, \nu = \pi/2$.}
\label{Fig.6}
\end{figure} 

We then generated a plot illustrating the correlation between the recoil energy of the nucleus and the scattering rate, considering the presence of mass splitting. Clearly, an increase in the mass splitting leads to higher values for both the position of the local maximum and the recoil energy. Figure \ref{Fig.6} displays the distribution on the $E_{R}-\delta_{\mathrm{DM}}$ plane. Across the entire plane distribution, an increase in $\delta_{\mathrm{DM}}$ leads to a higher value of $\bar{v}$. As a result, the phase space integral approaches its maximum value, allowing an earlier observation of the maximum local scattering rate at a low-energy position. When $\delta_{\mathrm{DM}}$ is negative, $\bar{v}$ decreases accordingly, indicating that a higher energy is required to reach the maximum local scattering rate. However, simultaneously, $v_{min}$ decreases, leading to an increased phase space integral. Consequently, we observe a higher abundance of dark matter at low-energy positions in the opposite direction of polarization. Remarkably, in contrast to elastic scattering, the event rate does not decrease as the recoil energy increases. This phenomenon can be attributed to the enhanced $\delta_{\mathrm{DM}}$.

For light dark matter with $m_{\chi}=1~\mathrm{GeV}$, information about recoiling electrons is obtained through the Migdal effect since the recoil of the nucleon cannot be observed. By integrating all recoil directions of the nucleon, our focus lies solely on the impact of the polarization direction angle $\nu$ and the electron recoil energy on the scattering rate for various dark matter mass splitting. 

%Analyzing equation (\ref{eqn:quadruple}), we observe from  Figure~\ref{Fig.3} $\nu-\delta_{\mathrm{DM}}$ that a significant number of cases occur only when the mass splitting is negative. In Figure~\ref{Fig.3}  $\nu-\delta_{\mathrm{DM}}$, while keeping other variables constant, equation (\ref{eqn:quadruple}) can be simplified to resemble the integration of $x^3\exp^{-x^2}$. Solving this integral reveals that when $x$ is maximized, it corresponds to the maximum value of the integral, which represents the highest number of cases. Here, variable $x$ corresponds to variable $E_R$ in equation (\ref{eqn:quadruple}), and considering that $E_R$ is a function of $\delta_{\mathrm{DM}}$ as equation (\ref{eqn: ER}). Based on the fact that $E_R \textgreater 0$, we can infer that when $\delta_{\mathrm{DM}}$ takes its minimum value, $E_R$ reaches its maximum value. Consequently, an evident number of cases are observed only in Figure $\nu-\delta_{\mathrm{DM}}$. Referring to equation (\ref{eqn: ER}), we observe that the order of $\mu v^2$ is identical to that of $\Delta$. Considering that energy must be positive, we can only observe the presence of mass splitting when $\Delta$ is negative, which indicates a heightened sensitivity to mass splitting. In the case of light dark matter and nucleon scattering, the recoil energy acquired by the nucleons is minimal, indicating that the recoil energy of the associated electrons is also relatively small when considering the Migdal effect.

\begin{figure}[!htbp]
    \centerline{ \includegraphics[width=9cm]{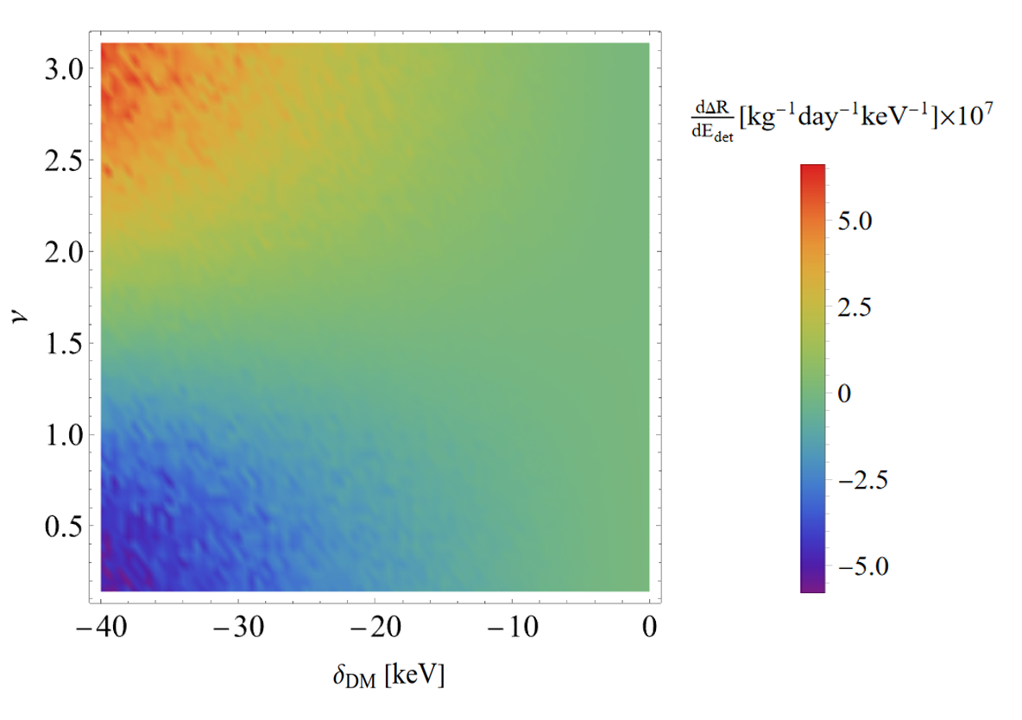}}
    \caption{Impact of mass splitting $\delta_{\text{DM}}$ and polarization angle $\nu$ on scattering probability in Migdal effect images, with electron recoil energy set at $E_{\text{det}} = 2~\mathrm{keV}$.}
    \label{Fig.7}
    \end{figure}

Similarly to heavy dark matter, the same distribution of dark matter can be observed in the polarization direction angle $\nu$, as depicted in Figure \ref{Fig.7}. Analyzing Figure~\ref{Fig.7}, we observe that the number of events is noticeable only when the mass splittings are negative. This is because $\Delta=\delta_{\text{DM}}+\delta_{\text{EM}}$, and $\delta_{\text{EM}}$ is non-zero under the Migdal effect due to $E_{\text{det}}$. Consequently, $\bar{v}\sim \Delta/q$ becomes significantly larger than $v_{\mathrm{esc}}$ in most parameter space. Therefore, the contribution of $\delta_{\text{EM}}$ must be counterbalanced by the negative $\delta_{\text{DM}}$ to achieve a smaller $\bar{v}$. Consequently, we only observe a number of cases when $\delta_{\text{DM}}$ is negative. In conclusion, the Migdal effect for polarized scattering is only available for negative mass splitting. 
 
    %\begin{figure}[H]
    %\centerline{ \includegraphics[width=10cm]{fig/figmgidalEdetδ.pdf}}
   %\caption{Correlation between electron recoil energy $E_{\text{det}}$ and scattering rate in the Migdal effect of dark matter with varying mass splitting. Special focus on angle variations: $\alpha = \pi/2$, $\beta = 0$, and nucleon polarization angle $\nu = \pi/2$.}
    %\label{Fig.7}
    %\end{figure}

\section{Conclusion}
\label{sec-3}

This study presents the novel application of spin-polarized direct detection to explore the inelastic signatures of dark matter, with a specific emphasis on the mass splitting and Migdal effect for heavy and light dark matter, respectively. Our comprehensive analyses unequivocally demonstrate the invaluable utility of the polarized triple-differential event rate as a powerful investigative tool for studying inelastic dark matter, owing to its ability to capture both angular and energy dependencies. Notably, in the case of pseudo-Dirac fermion dark matter, the angular dependence allows for the discrimination between positive and negative mass splitting, thereby offering a distinguishing characteristic. Moreover, an intriguing observation emerges, whereby the maxima of the event rate manifests at higher recoil energies $E_R$ as the mass splitting increases. These salient features strongly indicate that the angular and energy dependencies of the recoil rate provide complementary information for accurately determining mass splitting. Conversely, in the context of Migdal scattering, the event rate does not show a pronounced dependence on $\delta_{\mathrm{DM}}$. Only when $\delta_{\mathrm{DM}}$ is negative at specific $\nu$ values, a non-vanishing differential rate can be generated due to kinematic reasons. As a result, distinguishing between different mass splittings becomes feasible through a meticulous analysis of the event rate's shape.

\section*{Acknowledgement}
We would like to thank  Shu-Yuan Guo for the valuable discussions. This work is supported by the National Natural Science Foundation of China (NNSFC) under grant Nos. 12275232, 12005180, by the Natural Science Foundation of Shandong Province under Grant No. ZR2020QA083, and by the Project of Shandong Province Higher Educational Science and Technology Program under Grants No. 2022KJ271.

%\section{Appendix}

\appendix

\begin{widetext}

\section{Derivation of Equation ~(\ref{3.19}) }
\label{app:validation}
For the dark matter scattering process of ${\chi}_1N\longrightarrow{\chi}_2N$, the squared amplitude  $|\overline{M}|^2$ has 9 terms. The first term is
\begin{equation}
   \begin{aligned}
&\frac{1}{{m_{A^{\prime}}}^4} (256h_4^2\lambda_4^2m_N^2m^2\vec{S}_{\chi}^{ss^{\prime}}\cdot\vec{S}_{N}^{rr^{\prime}}\vec{S}_{\chi}^{s^{\prime}s}\cdot\vec{S}_{N}^{r^{\prime}r}
+256h_4^2\lambda_4^2m_N^2m\Delta\vec{S}_{\chi}^{ss^{\prime}}\cdot\vec{S}_{N}^{rr^{\prime}}\vec{S}_{\chi}^{s^{\prime}s}\cdot\vec{S}_{N}^{r^{\prime}r}
\\
&+64h_4^2\lambda_4^2m_N^2\Delta^2\vec{S}_{\chi}^{ss^{\prime}}\cdot\vec{S}_{N}^{rr^{\prime}}\vec{S}_{\chi}^{s^{\prime}s}\cdot\vec{S}_{N}^{r^{\prime}r})
\\
=&\frac{16m_N^2m^2}{{m_{A^{\prime}}}^4}(16h_4^2\lambda_4^2\vec{S}_{\chi}^{ss^{\prime}}\cdot\vec{S}_{N}^{rr^{\prime}}\vec{S}_{\chi}^{s^{\prime}s}\cdot\vec{S}_{N}^{r^{\prime}r}
+16h_4^2\lambda_4^2\frac{\Delta}{m}\vec{S}_{\chi}^{ss^{\prime}}\cdot\vec{S}_{N}^{rr^{\prime}}\vec{S}_{\chi}^{s^{\prime}s}\cdot\vec{S}_{N}^{r^{\prime}r})
\\
=&\frac{16m_N^2m^2}{{m_{A^{\prime}}}^4}(3h_4^2\lambda_4^2+3\frac{\Delta}{m}h_4^2\lambda_4^2), 
 \end{aligned}
\end{equation}
where $\Delta^2\simeq0$. The second term is

\begin{equation}
\begin{aligned}
&\frac{1}{{m_{A^{\prime}}}^4}4h_4^2\lambda_3^2
[\Delta\vec{K}\cdot\vec{S}_{N}^{rr^{\prime}}\delta^{ss^{\prime}}+4im_N\vec{S}_{\chi}^{ss^{\prime}}\cdot(\vec{S}_{N}^{rr^{\prime}}\times\vec{q})+4m_Nm\delta^{ss^{\prime}}\vec{v}^{\perp}\cdot\vec{S}_{N}^{rr^{\prime}}]
[\Delta\vec{K}\cdot\vec{S}_{N}^{r^{\prime}r}\delta^{s^{\prime}s}
\\
&+4im_N\vec{S}_{\chi}^{s^{\prime}s}\cdot(\vec{S}_{N}^{r^{\prime}r}\times\vec{q})-4m_Nm\delta^{s^{\prime}s}\vec{v}^{\perp}\cdot\vec{S}_{N}^{r^{\prime}r}]
\\
=&\frac{h_4^2\lambda_3^2}{{m_{A^{\prime}}}^4}
[64im_N^2\vec{S}_{\chi}^{ss^{\prime}}\cdot(\vec{S}_{N}^{rr^{\prime}}\times\vec{q})\vec{S}_{\chi}^{s^{\prime}s}\cdot(\vec{S}_{N}^{r^{\prime}r}\times\vec{q})
+64im_N^2m\delta^{ss^{\prime}}\vec{v}^{\perp}\cdot\vec{S}_{N}^{rr^{\prime}}\vec{S}_{\chi}^{s^{\prime}s}\cdot(\vec{S}_{N}^{r^{\prime}r}\times\vec{q})\\
&-64im_N^2m\vec{S}_{\chi}^{ss^{\prime}}\cdot(\vec{S}_{N}^{rr^{\prime}}\times\vec{q})\delta^{s^{\prime}s}\vec{v}^{\perp}\cdot\vec{S}_{N}^{r^{\prime}r}
-64m_N^2m^2\delta^{ss^{\prime}}\vec{v}^{\perp}\cdot\vec{S}_{N}^{rr^{\prime}}\delta^{s^{\prime}s}\vec{v}^{\perp}\cdot\vec{S}_{N}^{r^{\prime}r}
-4\Delta^2\vec{K}\cdot\vec{S}_{N}^{rr^{\prime}}\vec{K}\cdot\vec{S}_{N}^{r^{\prime}r}\delta^{s^{\prime}s}\delta^{ss^{\prime}}
\\
&+16im_N\Delta\vec{K}\cdot\vec{S}_{N}^{r^{\prime}r}\vec{S}_{\chi}^{ss^{\prime}}\cdot(\vec{S}_{N}^{rr^{\prime}}\times\vec{q})\delta^{s^{\prime}s}
-16im_N\Delta\vec{K}\cdot\vec{S}_{N}^{rr^{\prime}}\vec{S}_{\chi}^{s^{\prime}s}\cdot(\vec{S}_{N}^{r^{\prime}r}\times\vec{q})\delta^{ss^{\prime}}\\
&-16m_Nm\Delta\vec{v}^{\perp}\cdot\vec{S}_{N}^{r^{\prime}r}\vec{K}\cdot\vec{S}_{N}^{rr^{\prime}}\delta^{s^{\prime}s}\delta^{ss^{\prime}}
+16m_Nm\Delta\vec{v}^{\perp}\cdot\vec{S}_{N}^{rr^{\prime}}\vec{K}\cdot\vec{S}_{N}^{r^{\prime}r}\delta^{s^{\prime}s}\delta^{ss^{\prime}}] \\
=&0.
\end{aligned}
\end{equation}
The third item is
\begin{equation}
\begin{aligned}
&\frac{1}{{m_{A^{\prime}}}^4}16\lambda_3\lambda_4h_4^2m_N[-4im_N\Delta\vec{S}_{\chi}^{s^{\prime}s}\cdot(\vec{S}_{N}^{r^{\prime}r}\times\vec{q})\vec{S}_{\chi}^{ss^{\prime}}\cdot\vec{S}_{N}^{rr^{\prime}}
+4im_N\vec{S}_{\chi}^{ss^{\prime}}\cdot(\vec{S}_{N}^{rr^{\prime}}\times\vec{q})(2m\vec{S}_{\chi}^{s^{\prime}s}\cdot\vec{S}_{N}^{r^{\prime}r}+\Delta\vec{S}_{\chi}^{s^{\prime}s}\cdot\vec{S}_{N}^{r^{\prime}r})
\\
&+8m^2m_N\vec{v}^{\perp}\cdot\vec{S}_{N}^{rr^{\prime}}\vec{S}_{\chi}^{s^{\prime}s}\cdot\vec{S}_{N}^{r^{\prime}r}\delta^{ss^{\prime}}
+4mm_N\Delta\vec{v}^{\perp}\cdot\vec{S}_{N}^{rr^{\prime}}\vec{S}_{\chi}^{s^{\prime}s}\cdot\vec{S}_{N}^{r^{\prime}r}\delta^{ss^{\prime}}\\
&+4mm_N\Delta\vec{v}^{\perp}\cdot\vec{S}_{N}^{r^{\prime}r}\vec{S}_{\chi}^{ss^{\prime}}\cdot\vec{S}_{N}^{rr^{\prime}}\delta^{s^{\prime}s}
-2m\Delta\vec{K}\cdot\vec{S}_{N}^{rr^{\prime}}\vec{S}_{\chi}^{s^{\prime}s}\cdot\vec{S}_{N}^{r^{\prime}r}\delta^{ss^{\prime}}
-\Delta^2\vec{K}\cdot\vec{S}_{N}^{rr^{\prime}}\vec{S}_{\chi}^{s^{\prime}s}\cdot\vec{S}_{N}^{r^{\prime}r}\delta^{ss^{\prime}}
\\
&-\Delta^2\vec{K}\cdot\vec{S}_{N}^{r^{\prime}r}\vec{S}_{\chi}^{ss^{\prime}}\cdot\vec{S}_{N}^{rr^{\prime}}\delta^{s^{\prime}s}
-2m\vec{S}_{\chi}^{ss^{\prime}}\cdot\vec{S}_{N}^{rr^{\prime}}(4im_N\vec{S}_{\chi}^{s^{\prime}s}\cdot(\vec{S}_{N}^{r^{\prime}r}\times\vec{q})
-4mm_N\vec{v}^{\perp}\cdot\vec{S}_{N}^{r^{\prime}r}\delta^{s^{\prime}s}+\Delta\vec{K}\cdot\vec{S}_{N}^{r^{\prime}r}\delta^{s^{\prime}s}))]
\\
=&\frac{1}{{m_{A^{\prime}}}^4}\lambda_3\lambda_4h_4^2[\vec{S}_{\chi}^{s^{\prime}s}\cdot\vec{S}_{N}^{r^{\prime}r}(i128m_N^2m)\vec{S}_{\chi}^{ss^{\prime}}\cdot(\vec{S}_{N}^{rr^{\prime}}\times\vec{q})
+\vec{S}_{\chi}^{ss^{\prime}}\cdot\vec{S}_{N}^{rr^{\prime}}(-i128m_N^2m)\vec{S}_{\chi}^{s^{\prime}s}\cdot(\vec{S}_{N}^{r^{\prime}r}\times\vec{q})
\\
&+\vec{S}_{\chi}^{s^{\prime}s}\cdot\vec{S}_{N}^{r^{\prime}r}(i64m_N^2\Delta)\vec{S}_{\chi}^{ss^{\prime}}\cdot(\vec{S}_{N}^{rr^{\prime}}\times\vec{q})
+\vec{S}_{\chi}^{ss^{\prime}}\cdot\vec{S}_{N}^{rr^{\prime}}(-i64m_N^2\Delta)\vec{S}_{\chi}^{s^{\prime}s}\cdot(\vec{S}_{N}^{r^{\prime}r}\times\vec{q})]
\\
=&\frac{1}{{m_{A^{\prime}}}^4}16m_N^2m^2[-2\lambda_3\lambda_4h_4^2\vec{v}\cdot\vec{s}+2\lambda_3\lambda_4h_4^2\vec{v}^\prime\cdot\vec{s}
-\frac{\Delta}{m}\lambda_3\lambda_4h_4^2\vec{v}\cdot\vec{s}+\frac{\Delta}{m}\lambda_3\lambda_4h_4^2\vec{v}^\prime\cdot\vec{s}], 
\end{aligned}
\end{equation}
where $\vec{K}=-\vec{q}=-m(\vec{v}^\prime-\vec{v})$. 
The fourth item is
\begin{equation}
\begin{aligned}
&\frac{1}{{m_{A^{\prime}}}^4}4\lambda_4^2h_3^2(2(2m+\Delta)\vec{S}_{\chi}^{ss^{\prime}}\cdot(\vec{S}_{N}^{rr^{\prime}}\times\vec{q})
+i(4mm_N\vec{v}^{\perp}\cdot\vec{S}_{\chi}^{ss^{\prime}}\delta^{rr^{\prime}}-\Delta\vec{K}\cdot\vec{S}_{\chi}^{ss^{\prime}}\delta^{rr^{\prime}}))
\\
&\times (2(2m+\Delta)\vec{S}_{\chi}^{s^{\prime}s}\cdot(\vec{S}_{N}^{r^{\prime}r}\times\vec{q})
-i4mm_N\vec{v}^{\perp}\cdot\vec{S}_{\chi}^{s^{\prime}s}\delta^{r^{\prime}r}+i\Delta\vec{K}\cdot\vec{S}_{\chi}^{s^{\prime}s}\delta^{r^{\prime}r})
=0.
\end{aligned}
\end{equation}
The fifth item is
\begin{equation}
\begin{aligned}
&\frac{1}{{m_{A^{\prime}}}^4}\lambda_4^2h_3h_4[-128im^2m_N\vec{S}_{\chi}^{ss^{\prime}}\cdot\vec{S}_{N}^{rr^{\prime}}\vec{S}_{\chi}^{s^{\prime}s}\cdot(\vec{S}_{N}^{r^{\prime}r}\times\vec{q})
+128im^2m_N\vec{S}_{\chi}^{s^{\prime}s}\cdot\vec{S}_{N}^{r^{\prime}r}\vec{S}_{\chi}^{ss^{\prime}}\cdot(\vec{S}_{N}^{rr^{\prime}}\times\vec{q})
\\
&+128m_N^2m^2\vec{v}^{\perp}\cdot\vec{S}_{\chi}^{ss^{\prime}}\vec{S}_{\chi}^{ss^{\prime}}\cdot\vec{S}_{N}^{r^{\prime}r}\delta^{rr^{\prime}}
+128m_N^2m^2\vec{v}^{\perp}\cdot\vec{S}_{\chi}^{s^{\prime}s}\vec{S}_{\chi}^{s^{\prime}s}\cdot\vec{S}_{N}^{rr^{\prime}}\delta^{r^{\prime}r}
\\
&-32mm_N\Delta\vec{K}\cdot\vec{S}_{\chi}^{s^{\prime}s}\vec{S}_{\chi}^{ss^{\prime}}\cdot\vec{S}_{N}^{rr^{\prime}}\delta{r^{\prime}r}
+32mm_N\Delta\vec{K}.\vec{S}_{\chi}^{ss^{\prime}}\vec{S}_{\chi}^{s^{\prime}s}.\vec{S}_{N}^{r^{\prime}r}\delta{rr^{\prime}}
\\
&-16m_N\Delta^2\vec{K}\cdot\vec{S}_{\chi}^{s^{\prime}s}\vec{S}_{\chi}^{ss^{\prime}}\cdot\vec{S}_{N}^{rr^{\prime}}\delta{r^{\prime}r}
+16m_N\Delta^2\vec{K}\cdot\vec{S}_{\chi}^{ss^{\prime}}\vec{S}_{\chi}^{s^{\prime}s}\cdot\vec{S}_{N}^{r^{\prime}r}\delta{rr^{\prime}}
\\
&+32im_N\Delta^2\vec{S}_{\chi}^{s^{\prime}s}\cdot(\vec{S}_{N}^{r^{\prime}r}\times\vec{q})\vec{S}_{\chi}^{ss^{\prime}}\cdot\vec{S}_{N}^{rr^{\prime}}
-32im_N\Delta^2\vec{S}_{\chi}^{ss^{\prime}}\cdot(\vec{S}_{N}^{rr^{\prime}}\times\vec{q})\vec{S}_{\chi}^{s^{\prime}s}\cdot\vec{S}_{N}^{r^{\prime}r}
\\
&+64m_N^2m\Delta\vec{v}^{\perp}\cdot\vec{S}_{\chi}^{ss^{\prime}}\vec{S}_{\chi}^{s^{\prime}s}\cdot\vec{S}_{N}^{r^{\prime}r}\delta^{rr^{\prime}}
+64m_N^2m\Delta\vec{v}^{\perp}\cdot\vec{S}_{\chi}^{s^{\prime}s}\vec{S}_{\chi}^{ss^{\prime}}\cdot\vec{S}_{N}^{rr^{\prime}}\delta^{r^{\prime}r}
\\
&+128imm_N\Delta\vec{S}_{\chi}^{ss^{\prime}}\cdot(\vec{S}_{N}^{rr^{\prime}}\times\vec{q})\vec{S}_{\chi}^{s^{\prime}s}\cdot\vec{S}_{N}^{r^{\prime}r}
-128imm_N\Delta\vec{S}_{\chi}^{s^{\prime}s}\cdot(\vec{S}_{N}^{r^{\prime}r}\times\vec{q})\vec{S}_{\chi}^{ss^{\prime}}\cdot\vec{S}_{N}^{rr^{\prime}}]
\\
=&\frac{16\lambda_4^2h_3h_4m_N^2m^2}{{m_{A^{\prime}}}^4}[(2-3(1-\frac{m}{m_N})-\frac{\Delta}{2m}(1-\frac{m}{m_N})+\frac{\Delta}{2m_N})\vec{v}\cdot\vec{s}\\
&+(2-3(1+\frac{m}{m_N})-\frac{\Delta}{2m}(1+\frac{m}{m_N})-\frac{\Delta}{2m_N})\vec{v}^\prime\cdot\vec{s}].
\end{aligned}
\end{equation}
The sixth item is
\begin{equation}
\begin{aligned}
&\frac{1}{{m_{A^{\prime}}}^4}16\lambda_3^2h_3^2(m_N^2m^2\delta^{r^{\prime}r}\delta^{rr^{\prime}}\delta^{ss^{\prime}}\delta^{s^{\prime}s}
+16m_N^2m\Delta\delta^{r^{\prime}r}\delta^{rr^{\prime}}\delta^{ss^{\prime}}\delta^{s^{\prime}s}
+4m_N^2\Delta^2\delta^{r^{\prime}r}\delta^{rr^{\prime}}\delta^{ss^{\prime}}\delta^{s^{\prime}s})
\\
=&\frac{16m_N^2m^2\lambda_3^2h_3^2}{{m_{A^{\prime}}}^4}(1+\frac{\Delta}{m}).
\end{aligned}
\end{equation}
The seventh item is
\begin{equation}
\begin{aligned}
&\frac{1}{{m_{A^{\prime}}}^4}\lambda_3^2h_3h_4[-32imm_N^2\vec{S}_{\chi}^{s^{\prime}s}\cdot(\vec{S}_{N}^{r^{\prime}r}\times\vec{q})\delta^{rr^{\prime}}\delta^{ss^{\prime}}
+32imm_N^2\vec{S}_{\chi}^{ss^{\prime}}\cdot(\vec{S}_{N}^{rr^{\prime}}\times\vec{q})\delta^{r^{\prime}r}\delta^{s^{\prime}s}
\\
&+32m_N^2m^2\vec{v}^{\perp}\cdot\vec{S}_{N}^{r^{\prime}r}\delta^{rr^{\prime}}\delta^{ss^{\prime}}\delta^{s^{\prime}s}
+32m_N^2m^2\vec{v}^{\perp}\cdot\vec{S}_{N}^{rr^{\prime}}\delta^{r^{\prime}r}\delta^{s^{\prime}s}\delta^{ss^{\prime}}
\\
&-8m_Nm\Delta\vec{K}\cdot\vec{S}_{N}^{r^{\prime}r}\delta^{rr^{\prime}}\delta^{ss^{\prime}}\delta^{s^{\prime}s}
-8m_Nm\Delta\vec{K}\cdot\vec{S}_{N}^{rr^{\prime}}\delta^{r^{\prime}r}\delta^{s^{\prime}s}\delta^{ss^{\prime}}
\\
&-4m_N\Delta^2\vec{K}\cdot\vec{S}_{N}^{r^{\prime}r}\delta^{rr^{\prime}}\delta^{ss^{\prime}}
-4m_N\Delta^2\vec{K}\cdot\vec{S}_{N}^{rr^{\prime}}\delta^{r^{\prime}r}\delta^{s^{\prime}s}
\\
&-16im_N^2\Delta\vec{S}_{\chi}^{s^{\prime}s}\cdot(\vec{S}_{N}^{r^{\prime}r}\times\vec{q})\delta^{rr^{\prime}}\delta^{ss^{\prime}}
+16im_N^2\Delta\vec{S}_{\chi}^{ss^{\prime}}\cdot(\vec{S}_{N}^{rr^{\prime}}\times\vec{q})\delta^{r^{\prime}r}\delta^{s^{\prime}s}
\\
&+16mm_N^2\Delta\vec{v}^{\perp}\cdot\vec{S}_{N}^{r^{\prime}r}\delta^{rr^{\prime}}\delta^{ss^{\prime}}\delta^{s^{\prime}s}
+16mm_N^2\Delta\vec{v}^{\perp}\cdot\vec{S}_{N}^{rr^{\prime}}\delta^{r^{\prime}r}\delta^{s^{\prime}s}\delta^{ss^{\prime}}]
\\
=&\frac{16\lambda_3^2h_3h_4m_N^2m^2}{{m_{A^{\prime}}}^4}[(1-\frac{m}{m_N}+\frac{\Delta}{4m}(1-\frac{m}{m_N})-\frac{\Delta}{4m_N})\vec{v}\cdot\vec{s}
\\
&+(1+\frac{m}{m_N}+\frac{\Delta}{4m}(1+\frac{m}{m_N})+\frac{\Delta}{4m_N})\vec{v}^\prime\cdot\vec{s}].
\end{aligned}
\end{equation}
The eighth item is
\begin{equation}
\begin{aligned}
&\frac{1}{{m_{A^{\prime}}}^4}4\lambda_3\lambda_4h_3^2m_N(2i(2m+\Delta)\vec{S}_{\chi}^{s^{\prime}s}\cdot(\vec{S}_{N}^{r^{\prime}r}\times\vec{q})\delta^{rr^{\prime}}(2m\delta^{ss^{\prime}}+\Delta\delta^{ss^{\prime}})
\\
&+4mm_N\vec{v}^{\perp}\cdot\vec{S}_{\chi}^{s^{\prime}s}\delta^{rr^{\prime}}\delta^{r^{\prime}r}(2m\delta^{ss^{\prime}}+\Delta\delta^{ss^{\prime}})
+\delta^{r^{\prime}r}(-2i(2m+\Delta)\vec{S}_{\chi}^{ss^{\prime}}\cdot(\vec{S}_{N}^{rr^{\prime}}\times\vec{q})(2m\delta^{s^{\prime}s}+\Delta\delta^{s^{\prime}s})
\\
&+4mm_N\vec{v}^{\perp}\cdot\vec{S}_{\chi}^{ss^{\prime}}\delta^{rr^{\prime}}(2m\delta^{s^{\prime}s}+\Delta\delta^{s^{\prime}s})
\\
&-\delta^{rr^{\prime}}(\Delta\vec{K}\cdot\vec{S}_{\chi}^{s^{\prime}s}(2m\delta^{ss^{\prime}}+\Delta\delta^{ss^{\prime}})
+\Delta\vec{K}\cdot\vec{S}_{\chi}^{ss^{\prime}}(2m\delta^{s^{\prime}s}+\Delta\delta^{s^{\prime}s}))))
=0.
\end{aligned}
\end{equation}
The ninth item is
\begin{equation}
\begin{aligned}
&\frac{1}{{m_{A^{\prime}}}^4}4\lambda_3\lambda_4h_3h_4[16im^2m_N\vec{v}^{\perp}\cdot\vec{S}_{N}^{rr^{\prime}}\vec{S}_{\chi}^{s^{\prime}s}\cdot(\vec{S}_{N}^{r^{\prime}r}\times\vec{q})\delta^{ss^{\prime}}
+8imm_N\Delta\vec{v}^{\perp}\cdot\vec{S}_{N}^{rr^{\prime}}\vec{S}_{\chi}^{s^{\prime}s}\cdot(\vec{S}_{N}^{r^{\prime}r}\times\vec{q})\delta^{ss^{\prime}}
\\
&+16m^2m_N^2\vec{v}^{\perp}\cdot\vec{S}_{N}^{rr^{\prime}}\vec{v}^{\perp}\cdot\vec{S}_{\chi}^{s^{\prime}s}\delta^{ss^{\prime}}\delta^{r^{\prime}r}
+4im_N\Delta\vec{K}\cdot\vec{S}_{\chi}^{ss^{\prime}}\vec{S}_{\chi}^{s^{\prime}s}\cdot(\vec{S}_{N}^{r^{\prime}r}\times\vec{q})\delta^{rr^{\prime}}
\\
&+16m^2m_N^2\vec{S}_{\chi}^{s^{\prime}s}\cdot\vec{S}_{N}^{r^{\prime}r}\delta^{rr^{\prime}}\delta^{ss^{\prime}}+
8mm_N^2\Delta\vec{S}_{\chi}^{s^{\prime}s}\cdot\vec{S}_{N}^{r^{\prime}r}\delta^{rr^{\prime}}\delta^{ss^{\prime}}
\\
&-4mm_N\Delta\vec{v}^{\perp}\cdot\vec{S}_{N}^{r^{\prime}r}\vec{K}\cdot\vec{S}_{\chi}^{ss^{\prime}}\delta^{rr^{\prime}}\delta^{s^{\prime}s}
-4mm_N\Delta\vec{v}^{\perp}\cdot\vec{S}_{N}^{rr^{\prime}}\vec{K}\cdot\vec{S}_{\chi}^{s^{\prime}s}\delta^{r^{\prime}r}\delta^{ss^{\prime}}
\\
&+16m^2m_N^2\vec{S}_{\chi}^{ss^{\prime}}\cdot\vec{S}_{N}^{rr^{\prime}}\delta^{r^{\prime}r}\delta^{s^{\prime}s}
+8mm_N^2\Delta\vec{S}_{\chi}^{ss^{\prime}}\cdot\vec{S}_{N}^{rr^{\prime}}\delta^{r^{\prime}r}\delta^{s^{\prime}s}
\\
&-4im\Delta\vec{K}\cdot\vec{S}_{N}^{rr^{\prime}}\vec{S}_{\chi}^{s^{\prime}s}\cdot(\vec{S}_{N}^{r^{\prime}r}\times\vec{q})\delta^{ss^{\prime}}
-2i\Delta^2\vec{K}\cdot\vec{S}_{N}^{rr^{\prime}}\vec{S}_{\chi}^{s^{\prime}s}\cdot(\vec{S}_{N}^{r^{\prime}r}\times\vec{q})\delta^{ss^{\prime}}
\\
&-4mm_N\Delta\vec{v}^{\perp}\cdot\vec{S}_{\chi}^{s^{\prime}s}\vec{K}\cdot\vec{S}_{N}^{rr^{\prime}}\delta^{ss^{\prime}}
\delta^{r^{\prime}r}
+\Delta^2\vec{K}\cdot\vec{S}_{N}^{rr^{\prime}}\vec{K}\cdot\vec{S}_{\chi}^{s^{\prime}s}\delta^{ss^{\prime}}
\delta^{r^{\prime}r}
\\
&+\Delta^2\vec{K}\cdot\vec{S}_{N}^{r^{\prime}r}\vec{K}\cdot\vec{S}_{\chi}^{ss^{\prime}}\delta^{s^{\prime}s}
\delta^{rr^{\prime}}
+4mm_N\vec{v}^{\perp}\cdot\vec{S}_{\chi}^{ss^{\prime}}\delta^{rr^{\prime}}(-4im_N\vec{S}_{\chi}^{s^{\prime}s}\cdot(\vec{S}_{N}^{r^{\prime}r}\times\vec{q})
\\
&+4mm_N\vec{v}^{\perp}\cdot\vec{S}_{N}^{r^{\prime}r}\delta^{s^{\prime}s}-\Delta\vec{K}\cdot\vec{S}_{N}^{r^{\prime}r}\delta^{s^{\prime}s})
-2i\vec{S}_{\chi}^{ss^{\prime}}\cdot(\vec{S}_{N}^{rr^{\prime}}\times\vec{q})(-8(2m+\Delta)im_N\vec{S}_{\chi}^{s^{\prime}s}\cdot(\vec{S}_{N}^{r^{\prime}r}\times\vec{q})
\\
&-8mm_N^2\vec{v}^{\perp}\cdot\vec{S}_{\chi}^{s^{\prime}s}\delta^{r^{\prime}r}+8m^2m_N\vec{v}^{\perp}\cdot\vec{S}_{N}^{r^{\prime}r}\delta^{s^{\prime}s}
+4mm_N \Delta\vec{v}^{\perp}\cdot\vec{S}_{N}^{r^{\prime}r}\delta^{s^{\prime}s}
+2m_N\Delta\vec{K}\cdot\vec{S}_{\chi}^{s^{\prime}s}\delta^{r^{\prime}r}
\\
&-2m\Delta\vec{K}\cdot\vec{S}_{N}^{r^{\prime}r}\delta^{s^{\prime}s}-\Delta^2\vec{K}\cdot\vec{S}_{N}^{r^{\prime}r}\delta^{s^{\prime}s})
+8mm_N^2\Delta\vec{S}_{\chi}^{s^{\prime}s}\cdot\vec{S}_{N}^{r^{\prime}r}\delta^{ss^{\prime}}\delta^{rr^{\prime}}
+4m_N^2\Delta^2\vec{S}_{\chi}^{s^{\prime}s}\cdot\vec{S}_{N}^{r^{\prime}r}\delta^{ss^{\prime}}\delta^{rr^{\prime}}
\\
&+8mm_N^2\Delta\vec{S}_{\chi}^{ss^{\prime}}\cdot\vec{S}_{N}^{rr^{\prime}}\delta^{s^{\prime}s}\delta^{r^{\prime}r}+4m_N^2\Delta^2\vec{S}_{\chi}^{ss^{\prime}}\cdot\vec{S}_{N}^{rr^{\prime}}\delta^{s^{\prime}s}\delta^{r^{\prime}r}]=0.
\end{aligned}
\end{equation}

\end{widetext}

\bibliography{refs}

\end{document}